\documentclass[11pt,twoside]{article}

\usepackage{asp2004}
\usepackage{epsf}
\usepackage{lscape}

\begin{document}
\title{MASIV:The Microarcsecond Scintillation-Induced Variability Survey}   
\author{J.E.J. Lovell, D.L. Jauncey}   
\affil{Australia Telescope National Facility, PO Box 76, Epping NSW, Australia}
\author{C. Senkbeil}
\affil{School of Mathematics \& Physics, University of Tasmania, GPO Box 252 Hobart}
\author{S. Shabala}
\affil{Cavendish Laboratory, Cambridge CB3 OHE, United Kingdom}
\author{H.E. Bignall}
\affil{Joint Institute for VLBI in Europe, Postbus 2, 7990 AA Dwingeloo, The Netherlands}
\author{T. Pursimo}
\affil{Nordic Optical telescope, Sta Cruz de La Palma, E-38700 Tenerife, Spain}
\author{R. Ojha}
\affil{US Naval Observatory, 3450 Massachusetts Avenue NW, Washington DC 20392}
\author{J.-P. Macquart}
\affil{NRAO Jansky Fellow, Department of Astronomy, California Institute of Technology, Pasadena CA 91125}
\author{B.J. Rickett}
\affil{Department of Electrical and Computer Engineering, University of California, San Diego, La Jolla, CA 92093}
\author{M. Dutka}
\affil{US Naval Observatory, 3450 Massachusetts Avenue NW, Washington DC 20392}
\author{L. Kedziora-Chudczer}
\affil{9 School of Physics, University of Sydney, NSW 2006, Australia}

\begin{abstract} 
We are undertaking a large-scale 5 GHz VLA survey of the northern sky
to search for rapid intra-day variability (IDV). Over four observing
epochs we found that 56\% of the sources showed variability on
timescales of hours to several days. Fewer variables were seen at high
galactic latitudes, supporting interstellar-scintillation as the
principal mechanism responsible for IDV. We find evidence that many of
the scattering screens are not moving with the local standard of
rest. There are few scintillating sources seen at high redshift which
may be an indication of scattering in the turbulent intergalactic medium.
\end{abstract}

\section{Introduction}

Considerable evidence has now accumulated to demonstrate that
inter-stellar scintillation (ISS) in the turbulent, ionized
inter-stellar medium (ISM) of our Galaxy is the principal mechanism
responsible for the intra-day variability (IDV) seen in many
flat-spectrum AGN \citep[e.g.][and references therein]{big2006}. A
source must be small to scintillate; in the weak scattering case most
frequently observed at frequencies near 5 GHz, the source angular size
must be comparable to or smaller than the angular size of the first
Fresnel zone, which implies microarcsecond angular sizes for screen
distances of tens to hundreds of parsecs. The long time-scale over
which scintillations have been observed in some sources suggests that
such scintillating components are relatively long-lived despite their
small physical sizes. Interstellar scintillation probes AGN angular
sizes and brightness temperatures at cm-wavelengths that are
unachievable with VLBI.

\section{The MASIV Survey}

We are undertaking a large-scale 5 GHz VLA survey of a core sample of
525 compact, flat-spectrum AGN over the northern sky (declination $>$ 0
degrees) to search for IDV \citep{lov2003}. Our objective is to
assemble a statistically significant sample of sources exhibiting IDV
in order to address a number of important astrophysical questions.

The sources were chosen from the JVAS \citep{pat1992,bro1998,wil1998}
and CLASS \citep{mye1995} catalogues. Spectral indices were determined
from the JVAS and CLASS 8.5 GHz flux densities and the 1.4 GHz NVSS
catalogue \citep{con1998}, and a flat-spectrum sample selected to have
spectral indices greater than -0.3. We chose separate weak and strong
source samples of approximately 250 sources each. The weak sample
consists of sources with 8.5 GHz catalogued flux densities between 105
and 130 mJy, while the strong sample contains sources stronger than
600 mJy. Sources were selected to cover the northern sky with
individual source measurements being made every 2 hours, with 60
seconds on-source per scan. This gave $\sim$ 6 scans per source per
day, or approximately 10,000 scans per epoch. Primary flux density
calibration was based on observations with each sub-array of B1328+307
(3C286) and J2355+4950; the latter is a compact
gigahertz-peaked-spectrum source which is monitored at the VLA as part
of their regular calibrator-monitoring program.

The VLA was sub-divided into 5 sub-arrays of 5 or 6 antennas each, so
as to minimize slew times and hence maximize observing times. The
source list was subdivided into four declination bands and one
sub-array assigned to each. The fifth sub-array was used for more
intensive monitoring of selected sources known to show IDV. The
observations were undertaken during VLA reconfiguration times.

The observations took place over four 72 hour periods during January,
May and September 2002 (96h) and January 2003. This was done to ensure
that our survey sampled each source over the course of the year so
that sources that were in the ``slow'' part of their annual cycle would
not be missed. As a follow-up, we undertook a fifth three-day epoch of
MASIV observations in January 2006.

\section{Objectives}

Our principal objective was to produce a sample of at least 100 to 150
scintillators. Previous IDV surveys of the stronger flat-spectrum
sources had found roughly 15\% exhibited IDV; we started with a core
sample of 525 sources, the maximum number that we could observe once
every two hours sampled with the four sub-arrays. With such a large
sample size and number of scintillators, we can expect to explore
possible dependencies of the presence or absence of microarcsecond
angular structure in AGN on a variety of parameters.

A further objective was to survey a weaker sample to compare the
incidence of IDV in both the weak and strong sources. The inverse
Compton limit suggests that ISS may be more common amongst the weaker
sources since for a given angular size the brightness temperature
decreases with decreasing flux density.

We also hoped to find more of the very rapid scintillators like
PKS~0405--385 \citep{ked1997}, J1819+3845 \citep{den2000} and
PKS~1257--326 \citep{big2003}. As two of these three remarkable
sources were discovered serendipitously we hoped to find more such
sources and to ascertain how common they were and if they were more
likely amongst the weaker or stronger sources. Moreover, as such rapid
scintillation is due to scattering by nearby turbulent, ionized
clouds, the presence of new fast scintillators would signify the
presence of nearby clouds.

Only with a large sample of scintillators is it feasible to explore
the dependence of scintillators on parameters such as sky distribution
and hence to explore the structure of the turbulent ISM. Again a large
sample size allows us to search for any redshift dependence which
might signify the presence of turbulent, ionized material in the
intergalactic medium (IGM).

Microarcsecond angular resolution at cm wavelengths is not achievable
with Earth-based VLBI as it requires space baselines. Such long
baselines were not part of the VSOP Space VLBI mission \citep{hir1998}
and so interstellar scintillation remains the only way to achieve such
high angular resolution and to explore AGN structure and evolution at
the highest brightness temperatures at cm-wavelengths.

\section{Classification of Variables}

The uncertainties in the individual measurements are made up of two
components, a fractional error, p\%, due to pointing uncertainties,
and a noise error, s Jy, due to thermal noise and confusion. For the
present measurements we found that p = 1\% and s = 1.5 mJy.

As a first step towards finding the variables, sources were classified
as variable if their modulation index, defined as the rms of the three
days’ observations divided by the mean flux density, exceeded twice
that calculated from the above errors. On direct inspection of the
data, however, it soon became apparent that many of the slower
variables, that is sources with variability time-scales longer than
three days were not being noted as variable with the simple two-sigma
cut described above. We then tried a similar two-sigma selection but
this time based on the daily averages rather than the individual
measurements. This process yielded more variable detections, but again
inspection revealed that there remained many sources exhibiting
variability that were not being detected on either test.

The difficulty is that Chi-squared is not an ordered statistic, so
low-level, monotonic variability that is easily detected on inspection
was not being detected with these tests. We were left with no
alternative but to undertake ``by eye'' inspection and
classification. This was done independently by two of us. We adopted
the conservative null hypothesis that ``each source was considered
non-variable unless otherwise demonstrated''. The number of
``disagreements'' was small. Any source where there was disagreement on
its classification was reviewed and if we could not agree on the
classification, was classified non-variable. In practice most of those
sources where we disagreed were finally classified as non-variable.

Figure 1 shows some examples of the variability that we have uncovered
in the MASIV survey. J1819+3845 \citep{den2000} was
found to be by far the most dynamic variable with a modulation index
of 17\%, even though our 2-hourly sampling was at times less than the
variability time-scale. Variability was very clearly detected on each
of the four epochs. J0949+5819 demonstrates the sensitivity of our VLA
measurements even in our five-sub-array mode.
  
\begin{figure}[!ht]
\plottwo{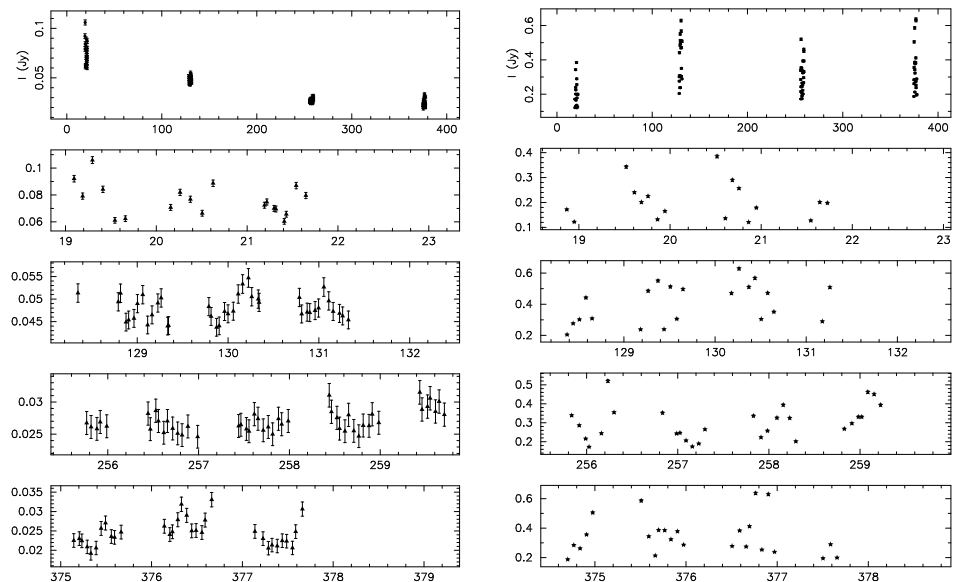}{lovell_j_fig1b.ps}
\caption{Examples of MASIV variability. J0949+5819 (left) and
  J1819+3845 (right). The top panel in each case shows all the total
  intensity data for the source against day number from 2002 January
  1. The lower four panels show light-curves for each of the first
  four epochs. The horizontal scale is the same (four days) in each
  case. }
\end{figure}

After removing 43 sources that showed structure or confusion, we were
left with a final sample of 482 sources. Figure 2 shows the
non-variable sources together with the numbers that were classified as
variable on one, two, three or all four epochs. 56 sources, 12\%, were
seen to vary on each of the four observing epochs.

\begin{figure}[!ht]
\plotone{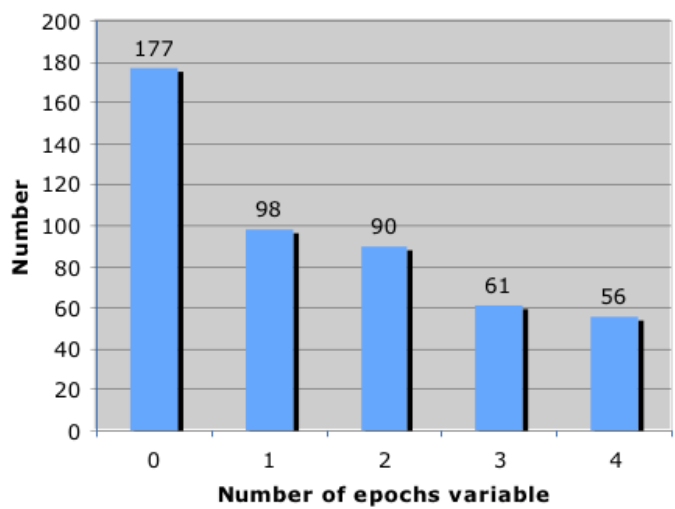}
\caption{The variability statistics for the 482 sources
  surveyed. Sources are divided into the number of epochs they were
  seen to vary. Sources not seen to vary in any epoch were classified
  non-variable, i.e. variable in zero epochs.}
\end{figure}

With any analysis of such a large number of observations, however,
false positives, i.e. sources that are incorrectly classified as
variable, are a significant concern. We believe that our
classification of the light curves as variable or not is reliable with
95\% confidence. The question then is how many false positives,
non-variables miss-classified, are there and how are they distributed?
Table 1 shows what the classifications would be if all 482 sources
were non-variable.

\begin{table}[!ht]
\caption{The probabilities and numbers of false variability
classifications were all 482 sources non-variable, together with the
observed numbers classified as variable.}
\smallskip
\begin{center}
\begin{tabular}{ccccc}
\tableline
\noalign{\smallskip}
Combination & Probability & Number of   & Fraction if none & Measured  \\
 &  & occurrences & are variable     & Numbers \\
\tableline
4 n      & $(1-P)^4$    & 1 & 81.5\%   & 177\\
3 n, 1 y & $P (1-P)^3$  & 4 & 17.2\%   & 98\\
2 n, 2 y & $P^2 (1-P)^2$ & 6 & 1.3\%    & 90\\
1 n, 3 y & $P^2 (1-P)^2$ & 4 & 0.04\%   & 61\\
4 y      & $P^4$        & 1 & 0.0006\% & 56\\
\tableline
\end{tabular}
\end{center}
\end{table}

The large number of sources classified as varying on multiple epochs
firmly establishes that our classification process is both accurate
and reliable. If all 482 sources were non-variable, then the expected
numbers exhibiting ``variability'' on 2, 3 or 4 epochs is a mere 6.7 or
1.4\%, compared with the observed 207 or 43\%. Moreover, the ratio of
false positives to non-variables is 17.4\%. Even if our classification
was 90\% reliable, then the number of false positives with 2 or more
detections remains less than 5\%.

This allows us to make a clear separation. We define as ``non-variable''
those 177 sources that showed no variability in any of the four
epochs, and we define as ``variable'' those 207 sources that showed
variability on two or more of the four epochs. With these definitions
we have two large and reliable samples each of approximately 200
sources, where the non-variables act as a control sample for the
variables. Each was drawn from the same selection criteria and cover
the same overall area of sky.

In the face of these false positives, an important question remains to
be answered, namely ``What fraction of the sources can be reliably
classified as variable?'' If N is the number of non-variables and T is
the number of one-time variables and F is the number of non-variables
classified as one-time variables, then T + F = 98 and N + T = 177 + 98
= 275, and F/N is, from Table 1, equal to 17.4\%. So that N = 214, F =
37 and T = 61, yielding the corrected total number of variables of
268. Therefore the fraction of variables to non-variables is 56\%, a
value which is significantly higher than found in any previous IDV
survey!

\section{Comparison of Variables and Non-Variables}

We have compared the fraction of strong and weak sources in our final
sample of 482 sources. There are close to equal fractions of
non-variables, one-time variables and 2+3+4 times variables in the
strong (37\%, 21\% and 42\%) and weak (37\%, 19\% and 44\%) parts of
our final sample. However, the two distributions differ in that the
weak sources contain a significantly higher fraction of 3-times and
4-times variables than the strong sample sources. \cite{lov2003} found
that the number of highly variable sources, those with rms flux
density variations greater than 4\% of the mean flux density,
increases with decreasing source flux density. Our finding an excess
of the 3-times and 4-times variables accounts well for this.

Those sources that were classified as varying on only one of the four
epochs form an interesting class of episodic variables. Correcting for
false positives leaves us with 61 sources, almost 13\% of the
total. Thus the microarcsecond components of AGN have life-times
lasting from less than four months, the time between our observational
epochs, to decades, the times over which some well-observed sources
have continued to show ISS, B0917+624 \citep{ric2001,jau2001},
PKS~1519--273 \citep{jau2003}, J1819+3845 \citep{den2003} and
PKS~1257--326 \citep{big2006}. It is only because of our sampling
over a full year that we find this unusual category of short-lived, episodic
scintillators.

A further surprise (at least to us) was the apparent absence of those
very rapid variables. J1819+3845 fell in our sample, but it was the
only source to show such variability. J0929+5013 showed rapid
variability in the January 2002 epoch \citep{lov2003} but,
although monitored closely, revealed only slower, many-hour
variability in the three later epochs. We had expected to find more of
these rapid variables especially given that two of the three known,
J1819+3845 and PKS1257-326 were discovered serendipitously.

One possibility is that such variability is predominantly episodic, as
is the case with PKS~0405--385 \citep{ked2001}. However, PKS~1257--326
and J1819+3845 are both long-lived over time-scales of up to a decade
and both were still scintillating when last observed. For PKS1257-326
the life-time is at least a decade \citep{big2006} and at least
half a decade for J1819+3845 \citep[and de
Bruyn private communication]{den2003}. For each of the three known fast
variables, it is the presence of the nearby scattering screen that is
responsible for the rapid variability. Thus it would seem that such
nearby screens cover very little of the sky.

\section{Variability Time-scales}

During the ``by eye'' examination we noted the presence of any
inflection points (i.e. change in sign of the derivative) at each
epoch in order to derive an indication of the variability time-scale
of each source. The majority of sources were found to show none or at
most one inflection point indicating variability timescales that are
predominantly longer than 3 days. The observed distribution of
inflection points is shown in Figure 3. Only a small number of sources
showed 2 or more inflection points. Overall, the distribution of
timescales was statistically the same for each epoch, remembering that
epoch 3 was four days rather than the three days of the other epochs.

\begin{figure}[!ht]
\plotone{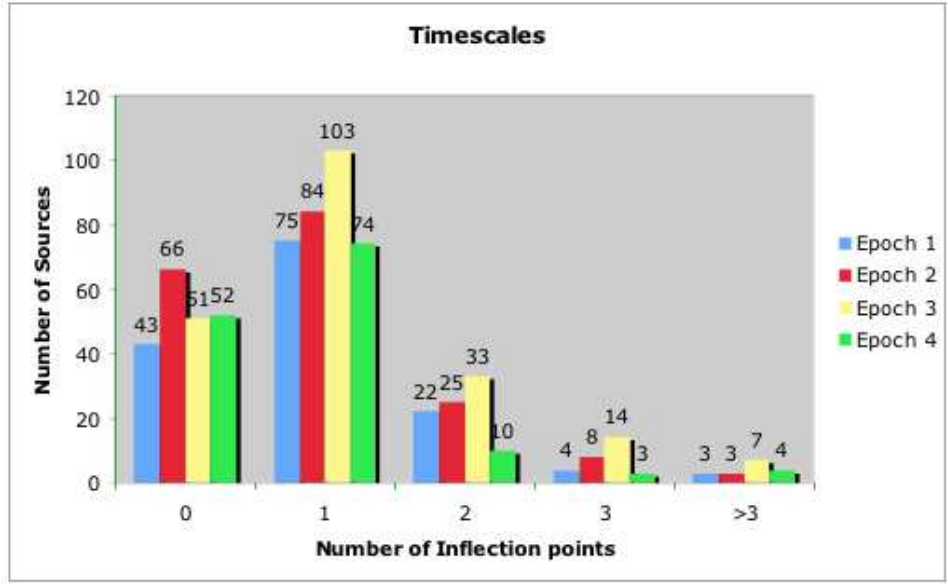}
\caption{The number of sources classified as variable verses the
  observed number of changes in the sign of the derivative of flux
  density vs time (i.e. number of inflection points). Clearly a
  majority of the sources vary on timescales of 3 days or longer.}
\end{figure}

A comparison of the distribution of inflection points for the weak and
strong sources revealed no significant difference between the two
classes. However $\sim 80$\% of all variables showed none or at most one
inflection point, indicating that we have only underestimates of
both the timescales and modulation indices for these sources.

The annual cycle seen in a number of sources is due to the changing
relative velocities of the Earth and the ISM responsible for the
scattering \citep{mac2003}. If the ISM velocities follow
the Local Standard of Rest, many sources would be expected to show a
slowing down in the third quarter of the year, and hence may more
easily be missed because of the lengthened time-scales. If the
scattering material responsible for the scintillation follows the
local standard of rest (LSR), then we would expect many more sources
to exhibit their slow-down during the September session. Figure 4
shows the numbers of variables found at each of the four epochs. A
contingency test shows no evidence that the numbers differ from the
mean in any epoch, even though epoch 3 lasted for four rather than
three days. The uniformity of variable numbers in each epoch suggests
a lack of evidence for a September slow-down, and it follows that the
majority of the scattering material is not moving at the LSR. This is
perhaps not unexpected; both PKS~1257--326 and J1819+3845, the two
sources for which reliable screen velocities have been measured, have
measured screen velocities that differ significantly from the LSR \citep{big2006,den2003}.

\begin{figure}[!ht]
\plotone{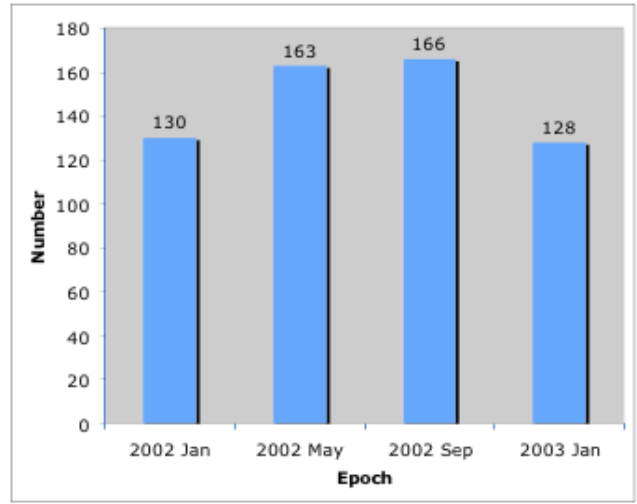}
\caption{Numbers of sources classified as variable in each
  epoch. Screens moving at the local standard of rest (LSR) would be
  expected to result in fewer variables beeing seen in epoch 3; we see
  no such deficit.}
\end{figure}

\section{Galactic Latitude Dependence}

A significant Galactic latitude dependence is to be expected in the
numbers of variable sources as inter-stellar scintillation is the
principal cause of this rapid intra-day variability. We have asked the
simple question ``are the latitude distributions of variables and
non-variables the same?'' A contingency test dividing them into two
samples, a low latitude sample, $|b| <$ 40 degrees, and high latitude
sample, $|b| >$ 40 degrees, shows that, at the 98\% confidence level,
the two distributions differ significantly. There are fractionally
more variables at low latitudes than there are at high latitudes, so
providing unambiguous support for the origin of the intra-day
variability being inter-stellar scintillation. This is in excellent
agreement with the modulation index dependence found by \cite{ric2006} in their analysis of the 146 flat-spectrum sources observed
with the Green Bank Interferometer.

\section{Spectral Index Dependence}

The MASIV sources were selected to have spectral indices greater than
-0.3 \citep{lov2003}. We have search for a spectral index
dependence by dividing the sources into two approximately equal groups
at a spectral index of 0.1 and testing if these fraction of
scintillators in these two groups are the same or not. The contingency
test tells us immediately that the probability is less than 1\% that
these two groups were drawn from the same population.

The spectral index distribution for sources surveyed at 5 GHz is
bi-modal with peaks near 0 and --0.7. These two peaks result from two
distinct source populations; the steep spectrum sources are
predominantly nearby galaxies, while the flat-spectrum sources are
predominantly distant quasars \citep[e.g.][]{con1974}. However,
the spectral index distribution of galaxies contains a significant
number with spectral indices flatter than --0.3, and it is these that
are likely responsible for the spectral index dependence noted here.

\section{Redshift Dependence}

We have extracted redshifts for 154 of the MASIV variables and
non-variables from the literature. In addition, we have measured
redshifts for a further 39 sources in a pilot program with the 2.5 m
Nordic Optical Telescope (Pursimo et al., in preparation) and the
overall redshift distribution is given in Figure 5. A contingency test
on the data in Figure 5 reveals that the redshift distributions of the
scintillators and non-scintillators are clearly different at the 97\%
confidence level, in the sense that there is a deficit of
scintillators amongst the high redshift sources.

We can test this distinction further by dividing the sources into two
separate groups, a ``near universe'', namely those with $z < 2$, and a
``far universe'', those with $z > 2$. Such a division is appropriate on
astrophysical and especially on observational grounds; those sources
with redshifts of 2 or more show the presence and pattern of the usual
strong quasar emission lines Lyman Alpha, Si IV, CIV and CIII], as
well as the presence of the Lyman Alpha absorption forest, which is
easily recognizable in establishing a redshift of 2 or more. Such a
division at $z = 2$ is thus extremely robust as the spectra of $z \geq 2$
quasars are easily and unambiguously recognizable, and avoids the
difficulties of an accurate redshift for quasars with only one
observed emission line, since they are clearly at $z < 2$ regardless of
the line identification. BL Lacs are also easily included as the
absence of any Lyman Alpha forest absorption classifies them as $z < 2$
as well.

With only two redshift categories, this maximizes the numbers and
provides a powerful statistical test which yields a formal probability
of less than 0.1\% that the low and high universe groups are drawn
from the same population. There is a highly significant deficit of
scintillators in the far universe.

We interpret the deficit of high redshift scintillators as due to
intergalactic scattering where interstellar scintillation acts as a
filter to distinguish sources containing apparent microarcsecond
components from those that do not. The absence of IDV in sources above
$z \sim 2$ indicates they are too large to scintillate. This implies that
sources at $z>2$ are either a) intrinsically too large to scintillate or
b) are being scatter-broadened by turbulence in the ionized
intergalactic medium at redshifts in excess of 2.
 
\begin{figure}[!ht]
\plotone{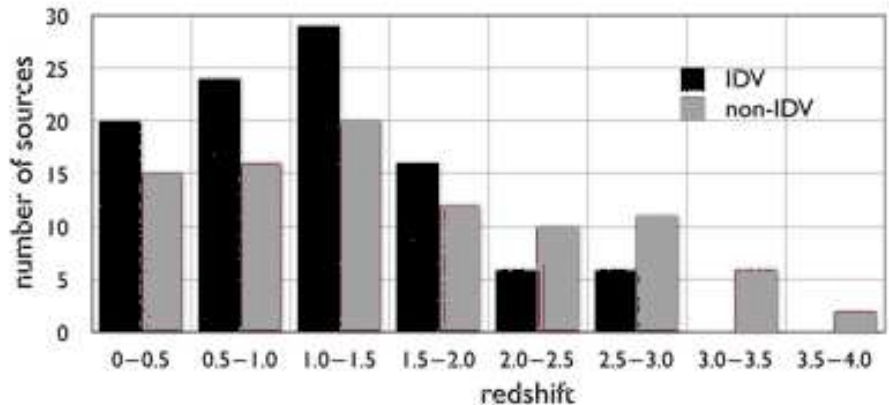}
\caption{The redshift distributions of the scintillators and
  non-scintillators. We detect a significant deficit of scintillators
  at redshifts above 2 which may be a signature of scattering in the
  turbulent ionized inter-galactic medium.}
\end{figure}

It seems unlikely that radio sources are intrinsically larger at high
redshift. However, we are concerned regarding the possibility that
selection effects present in the redshifts from the literature, which
constitute three quarters of the measured redshifts, may masquerade as
a redshift dependence. For example, more redshifts exist in the
literature for the stronger sources, so differences between the
stronger and weaker sources may show up as a redshift dependence. It
is for this reason that it is essential that we complete the redshift
measurements for all of the MASIV sources. Moreover, the importance of
a detection of the ionized, turbulent IGM argues that this be achieved
expeditiously.

\section{Conclusions}

\begin{itemize}
\item 56\% of sources varied on one or more of the four observing epochs
\item We found a significant Galactic latitude dependence with fewer variables at high Galactic latitudes. This strongly supports interstellar scintillation as the mechanism responsible for the observed variability.
\item Rapid, high modulation, inter-hour variable sources are rare
\item Number of variables in each epoch doesn't follow what is expected for LSR screens
\item We found a significant spectral index dependence, with fewer scintillators with smaller spectral indices.
\item Variability timescales are slow $\sim$1 day or more
\begin{itemize}
\item $\sim$10 ``fast'' sources (3 or more inflections)
\end{itemize}
\item We found few scintillators at redshifts of 2 and above. We argue that this is caused by scattering in the turbulent intergalactic medium.
\end{itemize}



\acknowledgements 

We are most grateful for the technical support provided by the NRAO staff in Socorro. The National Radio Astronomy Observatory is a facility of the National Science Foundation operated under a cooperative agreement by Associated Universities Inc. The Australia telescope national facility is funded by the Commonwealth of Australia for operation as a National facility managed by CSIRO.


\end{document}